\documentclass[sigconf]{acmart}
\AtBeginDocument{%
  \providecommand\BibTeX{{%
    \normalfont B\kern-0.5em{\scshape i\kern-0.25em b}\kern-0.8em\TeX}}}

\copyrightyear{2023}
\acmYear{2023}
\setcopyright{acmlicensed}\acmConference[SIGIR '23]{Proceedings of the 46th International ACM SIGIR Conference on Research and Development in Information Retrieval}{July 23--27, 2023}{Taipei, Taiwan}
\acmBooktitle{Proceedings of the 46th International ACM SIGIR Conference on Research and Development in Information Retrieval (SIGIR '23), July 23--27, 2023, Taipei, Taiwan}
\acmPrice{15.00}
\acmDOI{10.1145/3539618.3592002}
\acmISBN{978-1-4503-9408-6/23/07}

\usepackage{multirow}
\usepackage{rotating}

\begin{document}

\title{Improving Conversational Passage Re-ranking \protect\\ with View Ensemble}



\author{Jia-Huei Ju}
\affiliation{%
  \institution{Research Center for Information Technology Innovation, \\Academia Sinica}
  \country{}
}
\author{Sheng-Chieh Lin}
\affiliation{%
  \institution{David R. Cheriton School of Computer Science, \\University of Waterloo}
  \country{}
}
\author{Ming-Feng Tsai}
\affiliation{%
  \institution{Department of Computer Science, \\National Chengchi University}
  \country{}
}
\author{Chuan-Ju Wang}
\affiliation{%
  \institution{Research Center for Information Technology Innovation, \\Academia Sinica}
  \country{}
}

\renewcommand{\shortauthors}{Ju and Lin, et al.}

\begin{abstract}
This paper presents ConvRerank, a conversational passage re-ranker that employs a newly developed pseudo-labeling approach. Our proposed view-ensemble method enhances the quality of pseudo-labeled data, thus improving the fine-tuning of ConvRerank. Our experimental evaluation on benchmark datasets shows that combining ConvRerank with a conversational dense retriever in a cascaded manner achieves a good balance between effectiveness and efficiency. Compared to baseline methods, our cascaded pipeline demonstrates lower latency and higher top-ranking effectiveness. Furthermore, the in-depth analysis confirms the potential of our approach to improving the effectiveness of conversational search.
\end{abstract}






\maketitle

\section{Introduction}
\begin{figure}
    \centering
    \includegraphics[width=\linewidth]{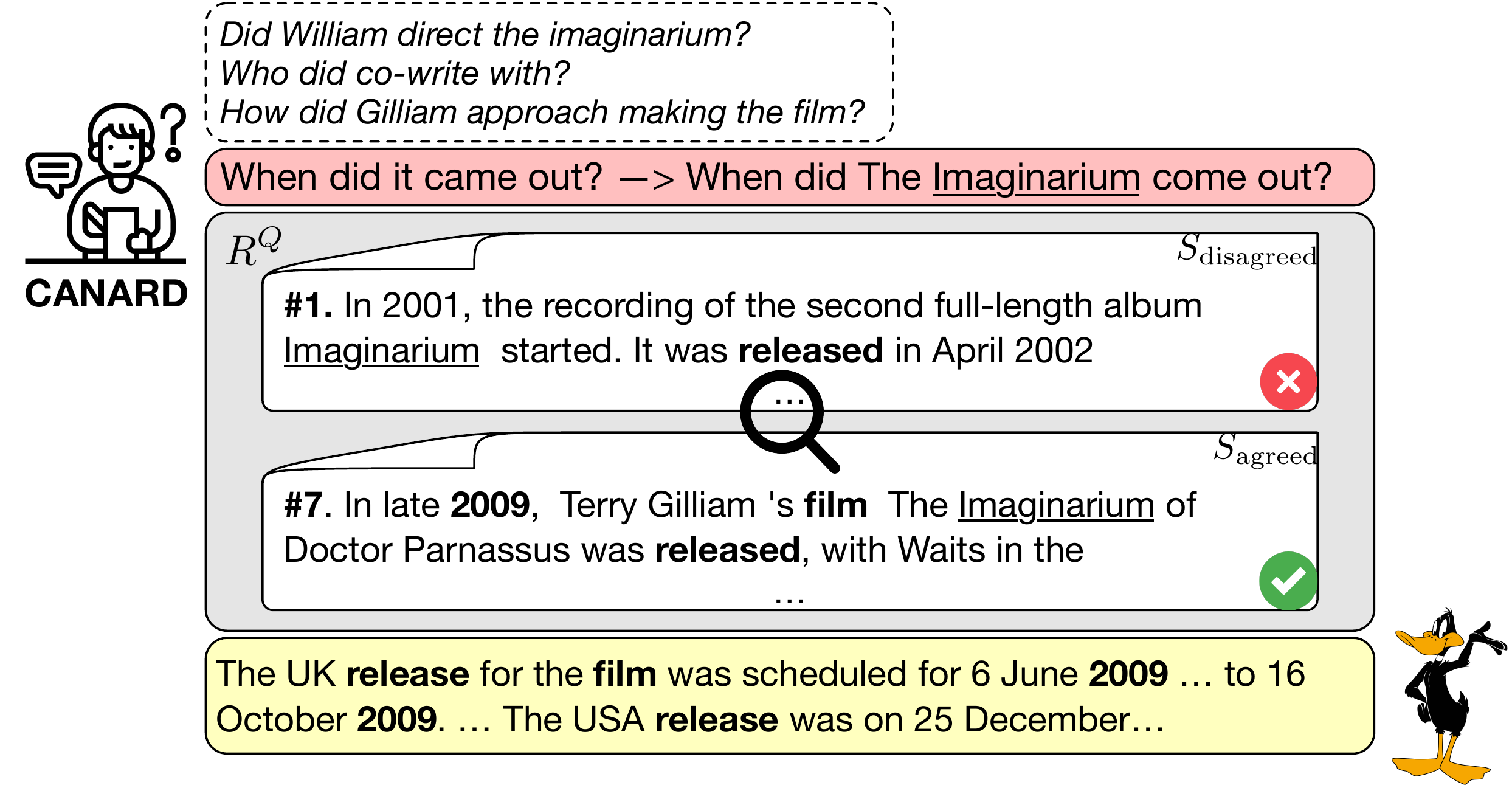}
    \caption{An example of the \emph{view-ensemble} method. $R^Q$ is an initial ranked list, with \#$k$ denoting the top-$k$ relevant passage. 
    Bold words indicate they appear in the ground-truth answer, the words underlined represent those in the question.}
    \label{fig:afpl}
\end{figure}
Conversational search (ConvSearch)~\cite{conv_search,SWIRL2018} has emerged as a rapidly growing research area as the popularity of conversational information seeking systems continues to rise.
ConvSearch has the potential to transform the way people search for information, moving from ad-hoc search to interactive search~\cite{zamani2022conversational, gao2018neural}.
However, the multi-turn nature of conversations poses significant challenges for information retrieval systems, as users often omit important contexts, particularly in the latter turns of conversations~\cite{choi2018quac, reddy2019coqa}.
This creates ambiguity in conversational queries, making it one of the most distinctive challenges for conversational AI systems~\cite{anantha2021open, reddy2019coqa, qu2020openretrieval, choi2018quac}. 
To facilitate research in this area, TREC has organized the Conversational Assistance Track (CAsT)~\cite{dalton2019trec, dalton2020trec} to create reliable benchmarks for the evaluation of ConvSearch systems.

Among all the ConvSearch systems, the multi-stage cascaded architecture has proven to be the most effective approach, which addresses the issue of query ambiguity in ConvSearch through the addition of a conversational query reformulation (CQR) module that employs
via 
heuristic~\cite{yang2019query,ilips} or neural approaches~\cite{lin2021multistage, voskarides2020query, yu2020few, vakulenko2021question, anantha2021open, kumar2020making}. 
Although effective, 
such additional modules may increase query latency and complexity, posing challenges for real-world deployment. 
Conversely, the recently proposed conversational dense retrieval (ConvDR)~\cite{yu2021few, lin2021contextualized, mao2022curriculum, krasakis2022zeroshot} greatly simplifies the ConvSearch system while demonstrating superior efficiency.
In ConvDR, a BERT-based query encoder~\cite{devlin2019bert, lan2020albert} 
encodes a user utterance and its dialogue context into a de-contextualized query embedding for dense retrieval without 
any query reformulation.
Despite its efficiency, ConvDR has shown to be less effective than state-of-the-art multi-stage ConvSearch systems~\cite{lin2021contextualized}, particularly in terms of top-ranking effectiveness (e.g., nDCG@3).

While the recent work~\cite{yu2021few, qian2022explicit} has integrated ConvDR into a cascaded architecture to improve effectiveness through passage re-ranking, we propose a more optimal re-ranker for ConvSearch which enhances effectiveness while reducing system complexity. With this in mind, we introduce a conversational passage re-ranker (ConvRerank)\footnote{Codes have been released at \url{https://github.com/cnclabs/codes.cs.sampling}.} that can be more seamlessly integrated with any first-stage retrieval methods for ConvSearch. 
The main advantages of our proposed ConvRerank over the previous works are two-fold: (1) Unlike \citet{qian2022explicit}, which relies on the separate query reformulation and passage re-ranking, ConvRerank is a single model capable of comprehending conversational queries and assessing their relevance to passages; (2) unlike the previous works of~\cite{lin2021contextualized, yu2021few} that used human-rewritten queries to construct pseudo-labeled training data, we propose a novel \emph{view-ensemble} pseudo-labeling approach that yields higher-quality data and facilitates fine-tuning of a more robust ConvRerank.

The example in Figure~\ref{fig:afpl} illustrates the motivation behind our view-ensemble method. In this example, a human reformulated the query by replacing ``it'' with ``Imaginarium.''
However, this reformulation is still ambiguous for search engines since ``Imaginarium'' can refer to either an album or a film.
As a result, the pseudo-relevance labels generated from previous approaches~\cite{lin2021contextualized, yu2021few} can be misleading (see \#$1$ in Figure~\ref{fig:afpl}). 
In this work, we recognize that the ground-truth answer (see the bottom block in Figure~\ref{fig:afpl}) for a given query contains useful words for clarifying information needs.
Based on this intuition, we explore how to better combine the information from a human-rewritten query and its answer and propose a simple yet effective view-ensemble method for generating training data with more accurate pseudo-relevance labels.

Our experiments on TREC CAsT~\cite{dalton2019trec, dalton2020trec} show that ConvRerank, a single model fine-tuned with our created pseudo labels, yields better re-ranking effectiveness than a more cumbersome re-ranking pipeline with a CQR and a re-ranking module. 
Furthermore, our in-depth analysis regarding the effects of different pseudo labels, first-stage retrieval, and model sizes confirms the robustness of the proposed method.
In addition, our ConvRerank can be integrated into any existing conversational retrieval methods, such as~\cite{yang2019query, lin2021contextualized}.

\section{Preliminaries}
\subsection{Task Definition and Notations}\label{sec:task}
The key distinction between ConvSearch and ad-hoc search lies in the interaction between queries.
While the latter utilizes a standard text query, ConvSearch utilizes a \emph{conversational query}, structured as a series of utterances. 
Formally, each conversational query, $q_i$, including the $i$-th turn utterance along with its 
conversational history (e.g., previous utterances), is defined as
$
    q_i = (u_i; u_1, u_2, ..., u_{i-1}).
$
Given a conversational query~$q_i$, the goal of ConvSearch systems is to retrieve a ranked list of relevant passages, denoted as $R = (p_1, p_2, \ldots, p_k)$.
The quality of $R$ can be measured using 
normalized discounted cumulative gain (nDCG).

\subsection{Cascaded Architecture for ConvSearch}\label{sec:cascaded}
\subsubsection{Conversational Query Reformulation}\label{sec:cqr}
ConvSearch systems have developed into complex pipelines. In particular, conversational query reformulation (CQR) is widely recognized as the most critical component of such systems~\cite{voskarides2020query, vakulenko2021question, anantha2021open, wu2022conqrr}.
The main goal of CQR is to transform a conversational query $q_i$ into an ad-hoc query $q'_i$, denoted as
$q'_i = \mathcal{F}_{\rm CQR} \big(u_i; u_1, u_2, ..., u_{i-1}\big)$.
As an example,
HQE~\cite{yang2019query} is 
a query expansion approach that appends context-dependant words extracted from historical conversations (i.e., $(u_1,u_2,\ldots,u_{i-1})$) to $u_{i}$ to form $q'_i$. 
Some researchers frame CQR as a term-classification task and utilize BERT models~\cite{devlin2019bert} to select tokens from historical conversations~\cite{voskarides2020query, kumar2020making}.
Moreover, some utilize transformer-based generative models~\cite{colin2020exploring, radford2019language} to rewrite queries through few-shot learning~\cite{yu2020few} or supervised learning~\cite{lin2020conversational, vakulenko2021question} on CANARD dataset~\cite{elgohary2019can}.

\subsubsection{Multi-stage Pipeline}\label{sec:multistage}
Many studies have adopted the standard multi-stage passage ranking pipeline in ad-hoc search~\cite{nogueira2019multistage} while using the reformulated query $q'$ as an ad-hoc query.
Examples include the works of~\cite{lin2021multistage,voskarides2020query, vakulenko2021question, kumar2020making}. Specifically, an effective ConvSearch system consists of a CQR module  $\mathcal{F}_{\rm CQR}$, 
a first-stage retriever $\mathcal{F}_{\rm RT}$ and a second-stage passage re-ranker $\mathcal{F}_{\rm RR}$, as follows: 
\begin{equation*}
    \mathcal{D}_{\rm RT} = 
    \mathcal{F}_{\rm RT}\big(q'; p\in \mathcal{D}\big),\; 
    R = 
    \mathcal{F}_{\rm RR}\big(q'; p\in \mathcal{D}_{\rm RT}\big),
\end{equation*}
where $\mathcal{D}$ denotes the entire passage collection, and $\mathcal{D}_{\rm RT}$ refers to a candidate passage set extracted by the first-stage retriever from $\mathcal{D}$ ($\vert\mathcal{D}\vert\gg \vert\mathcal{D}_{\rm RT}\vert$).
The passages in $\mathcal{D}_{\rm RT}$ are then sorted into a ranked list $R$ by the passage re-ranker.

\subsubsection{Dense Retrieval}\label{sec:dense}
Dense retrieval (DR) using a bi-encoder architecture with a passage encoder and a query encoder has gained attention for its effectiveness and efficiency in many knowledge-intensive tasks, as demonstrated in recent studies~\cite{reimers2019sentence, khattab2020colbert, karpukhin2020dense}.
DR works by precomputing representations of passages in a corpus through the passage encoder.
During retrieval, only the encoding of the query is performed, allowing for efficient end-to-end retrieval through inner product 
search~\cite{johnson2017computer}.
The bi-encoder architecture used in DR can be further optimized for conversational dense retrieval (ConvDR) by fine-tuning the model in a few-shot~\cite{yu2021few, mao2022curriculum, qian2022explicit} or weakly-supervised~\cite{lin2021contextualized} manner.
Despite its efficiency, ConvDR methods still fall short compared to multi-stage pipelines, particularly in terms of top-ranking effectiveness, as highlighted in~\cite{lin2021contextualized}. 

\section{Method}\label{sec:method}

\subsection{Pseudo-Labeling with Ensemble Views}\label{sec:method1}
Inspired by~\cite{lin2021contextualized}, we generate a ranked list for the 30K manually rewritten queries $q^*$ in the CANARD dataset~\cite{elgohary2019can}. 
We employ an effective two-stage retrieval pipeline~\cite{nogueira2019multistage} consisting of BM25 search and a monoT5~\cite{nogueira2020document} re-ranker to obtain the ranked list for $q^*$:
\begin{equation}
    R^Q =
    {\rm monoT5}\Big(q^*; p\in {\rm BM25}
    \big(q^*; p\in \mathcal{D}\big)\Big) 
    \label{eq:RQ},
\end{equation}
where $R^Q$ refers to the ranked list consisting of $M$ passages, which are re-ranked from the set of $N$ passage candidates via the BM25 retriever ($M<N$).\footnote{We follow previous works~\cite{lin2021contextualized} by setting $M=200$ and $N=1000$ in our experiments.}
Note that as in previous work by~\cite{lin2021contextualized}, we adopt the corpus $\mathcal{D}$ from CAsT~\cite{dalton2019trec}, which includes passages from TREC CAR~\cite{nanni2017benchmark} and MSMARCO~\cite{bajaj2016msmarco}. 

Motivated by previous works~\cite{kumar2020making,cormack2009reciprocal}, we propose further to leverage the \emph{answer} view and use these accurate signals to construct a ranked list with an \emph{ensemble} view.
First, to acquire the ranked list with the \emph{answer} view, we concatenate the query $q^*$ with the ground-truth answer $a$ from QuAC~\cite{choi2018quac}, which is an initial dataset of CANARD~\cite{elgohary2019can}.
We then pass it through the same retrieval pipeline as in Eq.~\eqref{eq:RQ}, obtaining the answer-view ranked list
\begin{equation}
    R^A = {\rm monoT5}\Big(q^*; p\in {\rm BM25}\big(q^* \mathbin\Vert a; p\in \mathcal{D}\big)\Big)\label{eq:RA},
\end{equation}
where $\mathbin\Vert$ denotes the concatenation operator.
With the two ranked lists (i.e., $R^Q$ and $R^A$), we define a filtering function $\Phi$ to generate a ranked list with \emph{ensemble} views as
\begin{align}
    R^{{\rm EM}(R^Q|R^A)} = \Phi(R^Q,R^A)=S_{\rm agreed} \mathbin\Vert S_{\rm disagreed},
    \label{eq:RQA} 
\end{align}
where $R^{{\rm EM}(R^Q|R^A)}$ denotes the ranked list with an ensemble view that $R^A$ serves as a filter towards $R^Q$, consisted of two ordered lists:
\begin{align*}
    S_{\rm agreed}=&\big(p_1^+,p_2^+,\ldots,p_{\ell}^+\big), \\
    S_{\rm disagreed} =&\big(p_1^-,p_2^-,\ldots,p_h^-\big),
    \notag
\end{align*}
where $p_i^+$ denotes the passage agreed by both views (i.e., $p_i^+$ in both $R^Q$ and $R^A$), and $p_j^{-}$ denotes the passage in $R^Q$ but not in $R^A$.
Note that we here keep the original relative order of passages in $R^Q$ for the aforementioned two ordered lists.

In other words, the function $\Phi$ reorders the passages in $R^Q$ by pushing the passages agreed by both $R^Q$ and $R^A$ forward and moves the ones only in $R^Q$ backward.
The motivation behind this design is that a stand-alone query is often ambiguous for search engines~\cite{song2007identifyig};
This ambiguity is even more critical in the context of ConvSearch (See Figure~\ref{fig:afpl}).
As a result, relying solely on $R^Q$ to synthesize pseudo relevance for model training may cause re-rankers to establish unfaithful relations between passages and conversational context.
To address this issue, we combine the ranked list with the answer view to reorder passages in $R^Q$; that is, 1) the resulting passages in $S_{\rm agreed}$ should be more aligned with the user's information need, and 2) the ones in $S_{\rm disagreed}$ could serve as hard negative to facilitate a more effective training of ConvRerank.

\subsection{Fine-tuning Conversational Passage re-rankers with Pseudo-Labeling} \label{sec:ConvRerank}
For training the proposed ConvRerank, 
we adopt the ensemble-view ranked list $R^{{\rm EM}(R^Q|R^A)}$ to synthesize pseudo relevance.
Specifically, for each query, we generate the pseudo labels by treating the top-$k$ results in the ensemble-view ranked list as (pseudo) positive labels and randomly sampling $k$ passages from the top-$k$ to $M$ passages in the same list as (pseudo) negative labels.\footnote{Note that we set $k$ to 40, which is found to be optimal in our experiments}
As for the backbone architecture of ConvRerank, we use T5 models~\cite{colin2020exploring} and recast the input format of conversational query passage pairs $(q_i=(u_i; u_1, u_2, ..., u_{i-1}), p)$ as a text-to-text format: 
\begin{equation*}
    \resizebox{\hsize}{!}{$
    \texttt{Query: } u_i 
    \texttt{ Context: } \Omega(u_1,u_2,\ldots, u_{i-1})
    \texttt{ Document: } p
    \texttt{ Relevant:}$}
\end{equation*}
where $\Omega$ indicates the join function with a special unused token in T5 vocabulary ``\texttt{<extra\_id\_10>}'' as a separation token between each element (i.e., each historical utterance).
The objective is the negative log-likelihood loss of generating \texttt{true/false} tokens for relevant/irrelevant passages. 
We compute the relevance scores by taking the probability of \texttt{true/false} logit values following the approach of monoT5~\cite{nogueira2020document}.
It is worth noting that while our focus is on re-ranking and ConvSearch, the proposed pseudo-labeling method can be applied to ConvDR and other IR tasks as well.

\section{Experiments}
\begin{table}
    \centering
    \caption{TREC CAsT statistics.}
    \vspace{-0.4cm}
    \label{tab:cast-stats}
    \resizebox{.28\textwidth}{!}{
    \begin{tabular}{lrr}
    \toprule
         & CAsT'19 Eval & CAsT'20 Eval \\
    \midrule
         \# Queries     & 173 & 208    \\
         \# Topics      & 20  & 25     \\
         \# Judgements  & 29,571 & 40,451 \\
    \midrule
         \# Passages    & \multicolumn{2}{c}{38M} \\
    \bottomrule 
    \end{tabular}
    }
\end{table}

\subsection{Data and Experimental Setups}
\subsubsection{TREC CAsT Evaluation Topics.}
We used benchmark evaluation data from the TREC Conversational Assistant Track (CAsT): CAsT'19 Eval~\cite{dalton2020trec} and CAsT'20 Eval~\cite{dalton2019trec}.
Each data includes TREC-judged topics; each topic has approximately 8 to 10 turns of questions, and
the relevance judgment adopts a five-point scale from~0 to~4. The corpora are composed of MSMARCO~\cite{bajaj2016msmarco} and TREC CAR~\cite{nanni2017benchmark}. 
The data statistics are presented in Table~\ref{tab:cast-stats}.

\subsubsection{Training, Inference, and Evaluation.}
We first initialized our ConvRerank with the monoT5~\cite{nogueira2020document} checkpoint, a T5-base re-ranking model that has been fine-tuned on MSMARCO~\cite{bajaj2016msmarco}.\footnote{We found that fine-tuning from scratch yields a significant effectiveness drop.}
We then fine-tuned the model using our synthesized pseudo labels (see Section~\ref{sec:method}) with the batch size of 256 for 5 epochs, which is chosen based on the performance on the CAsT'19 train set, within the range of 1 to 5. 
The other settings for fine-tuning, such as the learning rate and sequence length, are the same as monoT5~\cite{nogueira2020document}.
We re-rank the top 100 passage candidates retrieved from CQE~\cite{lin2021contextualized} and compare their top-ranking and overall effectiveness, as measured by nDCG@3 and @100, respectively. 
The latency of the re-ranking stage\footnote{During re-ranking, we set the maximum token length for each document to 384 and the remaining 128 for the query and its context.}
was measured on Google Colab with an A100 GPU. 

\begin{table}
    \centering
    \caption{Evaluation on CAsT datasets. ‘$\dagger$’ indicates top-500 passage re-ranking; the other systems use top-100 passages. Score with superscript indicates it greater ($p\leq0.05$) than those one superscripted letters on paired $t$-tests.}
    \vspace{-0.4cm}
    \label{tab:main-a}
    \resizebox{.49\textwidth}{!}{
    \begin{tabular}{l l r cc cc}
    \toprule
         & & Latency &
         \multicolumn{2}{l}{CAsT'19 Eval} & 
         \multicolumn{2}{l}{CAsT'20 Eval } \\
         \cmidrule(lr){3-3}
         \cmidrule(lr){4-5}
         \cmidrule(lr){6-7}
         \# & Retrieval ($\rightarrow$ Re-ranking) & (ms/q) &
         \multicolumn{2}{l}{nDCG@3 / 100} &
         \multicolumn{2}{l}{nDCG@3 / 100} \\
    \midrule  
         & \multicolumn{3}{l}{{\bf Upper-bound system w/ manual query}} \\
         & TCT-ColBERT~\cite{lin2021inbatch} $\rightarrow$ monoT5
         & - &  
        \multicolumn{2}{l}{0.583 / 0.545} & 
        \multicolumn{2}{l}{0.556 / 0.546} \\
    \midrule
         (a) &ConvDR $\rightarrow$ BERT (RRF)~\cite{yu2021few} 
         & 1900 & 
         \multicolumn{2}{l}{0.541 / -} &
         \multicolumn{2}{l}{0.392 / -} \\
         (b) &CRDR~\cite{qian2022explicit} 
         & 1690 & 
         \multicolumn{2}{l}{0.553 / - } &
         \multicolumn{2}{l}{0.381 / - }\\
         (c) &CTS+MVR$^\dagger$~\cite{kumar2020making} 
         & 14630 & 
         \multicolumn{2}{l}{{\bf 0.565} / - } &
         \multicolumn{2}{l}{- / -} \\
    \midrule  
         (d) & CQE 
         & - & 
         \multicolumn{2}{l}{0.492 / 0.447} & 
         \multicolumn{2}{l}{0.319 / 0.350}\\
         (e) & CQE $\rightarrow$ T5-rewrite+monoT5
         & 1910 &
         \multicolumn{2}{l}{0.549$^{d}$ / 0.484$^{d}$ } & 
         \multicolumn{2}{l}{0.418$^{d}$ / 0.395$^{d}$ }\\
         (f) & CQE $\rightarrow$ ConvRerank
         & 1675 &
         \multicolumn{2}{l}{{\bf 0.563}$^{d}$ / {\bf 0.487}$^{d}$} & 
         \multicolumn{2}{l}{{\bf 0.432}$^{d}$ / {\bf 0.456}$^{de}$}\\
    \bottomrule
    \end{tabular}
    }
\end{table}

\subsection{Experimental Results}

Table~\ref{tab:main-a} presents our experimental results. 
First, in the second block of the table, we compared our cascaded approach (i.e., (f) CQE $\rightarrow$ ConvRerank) to other multi-stage systems, including
(a) ConvDR $\rightarrow$ BERT (RRF)~\cite{yu2021few}, which is a rank fusion~\cite{cormack2009reciprocal} of few-shot ConvDR and BERT re-ranker, 
(b) CRDR~\cite{qian2022explicit}, which integrates ConvDR and a query modification module for further BERT re-ranking, and 
(c) CTS+MVR~\cite{kumar2020making}, which utilizes multiple query views and BERT-base re-ranking to fuse over the views.
We observe that our approach yields better efficiency and effectiveness (especially in CAsT'20) compared to these systems.
This result demonstrates the advantages of ConvRerank over the other re-ranking solutions for conversational search. 
Second, we compare the passage re-ranking effectiveness of ConvRerank with the baseline re-rankers: the monoT5 reranker~\cite{nogueira2020document} with a T5-base query rewriting model~\cite{lin2020conversational}.
As shown in the last panel of Table~\ref{tab:main-a}, ConvRerank outperforms the baseline re-ranker, monoT5 with T5-rewrite, on all evaluation sets.
In terms of efficiency, ConvRerank, which does not require conversational query rewriting, achieves lower overall latency compared to monoT5 with T5-rewrite, making it a more efficient option.

Note that compared to CAST'19, 
CAsT'20 requires more complex conversational query understanding from user utterances and system responses~\cite{dalton2019trec}; thus, the larger gap in CAsT'20 between our system and the others indicates that ConvRerank can address more challenging conversational queries. 
However, all the systems still lag behind the one using human-rewritten queries (the first row), indicating there is still room for improvement for future research.

\subsection{Effect Analysis}\label{sec:effect}

\subsubsection{Pseudo Labels}\label{sec:effect-filtering}
To examine the effect of pseudo labels for fine-tuning ConvRerank, Table~\ref{tab:main-b} compares the effectiveness of models trained on the data with pseudo labels from different ranked lists: (a) $R^{\rm EM (R^Q|R^A)}$, our approach; (b) $R^Q$ in Eq.~\eqref{eq:RQ}; (c) $R^A$ in Eq.~\eqref{eq:RA}; (d) $R^{\rm EM (R^A|R^Q)}$, another ranked list also with the {ensemble} view by reversing the two lists in Eq.~\eqref{eq:RQA}.
We observe that the re-rankers trained on the pseudo labels generated from the ranked lists with the \emph{ensemble} view (i.e., (a) and (d)) outperform their corresponding single-view variants. 
(i.e., (b) and (c)) 
This result demonstrates that $R^Q$ and $R^A$ provide different views for conversational search and can complement each other well.
It is worth noting that human-reformulated queries alone ($R^Q$) generate better training data than those combined with answers ($R^A$).

\begin{table}
    \centering
    \caption{Fine-tune with different pseudo-labels. Score with superscript is greater than ($p\leq 0.05$) those superscripted.}
    \vspace{-0.4cm}
    \label{tab:main-b}
    \resizebox{.42\textwidth}{!}{
    \begin{tabular}{ll ll ll}
    \toprule
         & &
         \multicolumn{2}{l}{CAsT'19 Eval} &
         \multicolumn{2}{l}{CAsT'20 Eval} \\
         \cmidrule(lr){3-4}
         \cmidrule(lr){5-6}
         \# & Ranked list & 
         \multicolumn{2}{l}{nDCG@3 / 100} &
         \multicolumn{2}{l}{nDCG@3 / 100} \\
    \midrule
         (a) & $R^{{\rm EM}(R^Q | R^A)}$ (proposed) &
         \multicolumn{2}{l}{{\bf 0.563}$^{bcd}$ / {\bf 0.487}$^{bcd}$} & 
         \multicolumn{2}{l}{{\bf 0.432}$^{bcd}$ / {\bf 0.456}$^{bcd}$} \\
         (b) & $R^Q$ &
         \multicolumn{2}{l}{0.517 / 0.467} & 
         \multicolumn{2}{l}{0.396 / 0.382}\\
         (c) & $R^A$ &
         \multicolumn{2}{l}{0.495 / 0.464} & 
         \multicolumn{2}{l}{0.392 / 0.382} \\
         (d) & $R^{{\rm EM}(R^A | R^Q)}$ &
         \multicolumn{2}{l}{0.519$^{c}$ / 0.474$^{bc}$} &
         \multicolumn{2}{l}{0.403 / 0.389$^{bc}$} \\
    \bottomrule
    \end{tabular}
    }
\end{table}

\begin{table}[]
    \centering
    \caption{Evaluation on different first-stage retrieval. ‘$\ddagger$’ indicates top-1000 passage re-ranking; the others use top-100.}
    \vspace{-0.4cm}
    \resizebox{.45\textwidth}{!}{
    \begin{tabular}{clllll}
    \toprule
         &  &
         \multicolumn{2}{l}{CAsT'19 Eval} & 
         \multicolumn{2}{l}{CAsT'20 Eval} \\
         \cmidrule(lr){3-4}
         \cmidrule(lr){5-6}
         & Retrieval ($\rightarrow$ Re-ranking) & 
         \multicolumn{2}{l}{nDCG@3 / 100} & 
         \multicolumn{2}{l}{nDCG@3 / 100} \\
    \midrule
         \multirow{3}{*}{\begin{turn}{90}Sparse\end{turn}} &
         HQE~\cite{yang2019query} &
         \multicolumn{2}{l}{0.261 / 0.308} & 
         \multicolumn{2}{l}{0.164 / 0.204} \\
         &
         HQE $\rightarrow$ T5-rewrite + monoT5$^\ddagger$& 
         \multicolumn{2}{l}{0.553 / {\bf 0.519}} & 
         \multicolumn{2}{l}{0.379 / 0.377} \\
         &
         HQE $\rightarrow$ ConvRerank$^\ddagger$ & 
         \multicolumn{2}{l}{{\bf 0.558} / 0.511} &
         \multicolumn{2}{l}{{\bf 0.389} / {\bf 0.384}} \\
    \midrule
         \multirow{3}{*}{\begin{turn}{90}Dense\end{turn}} &
         CQE~\cite{lin2021contextualized} &
         \multicolumn{2}{l}{0.492 / 0.447} & 
         \multicolumn{2}{l}{0.319 / 0.350} \\
         &
         CQE $\rightarrow$ T5-rewrite + monoT5& 
         \multicolumn{2}{l}{0.549 / 0.484} & 
         \multicolumn{2}{l}{0.418 / 0.395} \\
         &
         CQE $\rightarrow$ ConvRerank &  
         \multicolumn{2}{l}{{\bf 0.563} / {\bf 0.487}} & 
         \multicolumn{2}{l}{{\bf 0.432} / {\bf 0.456}} \\
     \midrule
         \multirow{3}{*}{\begin{turn}{90}Hybrid\end{turn}} &
         CQE-{\small HYB}~\cite{lin2021contextualized} &
         \multicolumn{2}{l}{0.498 / 0.494} & 
         \multicolumn{2}{l}{0.330 / 0.368} \\
         &
         CQE-{\small HYB} $\rightarrow$ T5-rewrite + monoT5& 
         \multicolumn{2}{l}{0.556 / 0.531} & 
         \multicolumn{2}{l}{{\bf 0.428} / {\bf 0.411}} \\
         &
         CQE-{\small HYB} $\rightarrow$ ConvRerank & 
         \multicolumn{2}{l}{{\bf 0.584} / {\bf 0.534}} & 
         \multicolumn{2}{l}{0.424 / 0.410} \\
    \bottomrule
    \end{tabular}
    }
    \label{tab:ablation-a}
\end{table}

\subsubsection{First-stage Retrievers.}
To examine the robustness of ConvRerank, we evaluated its performance with two other first-stage retrieval methods: (1) HQE~\cite{yang2019query}, a sparse retriever, built upon BM25 that heuristically concatenates words from the historical conversation;
(2) CQE-{\small HYB}~\cite{lin2021contextualized}, a hybrid retriever that combines CQE and CQE-sparse.\footnote{CQE-sparse is a variant of CQE that employs $L_2$-norm to select words from the historical context as query expansion for BM25 search.}
Note that neither of the two approaches requires neural models for reformulating conversational queries, which is consistent with our goal of building a simple yet effective system.
Table~\ref{tab:ablation-a} tabulates the performance with different first-stage retrieval, including the originally adopted CQE and aforementioned approaches.
We observe that ConvRerank is able to yield improvement upon different first-stage retrieval methods.
Notably, ConvRerank works effectively with HQE and sometimes performs on par with dense retrieval approaches; for example, on CAsT'19, HQE $\rightarrow$ ConvRerank achieves a similar nDCG@3 score to CQE-{\small HYB} $\rightarrow$ monoT5 (i.e., 0.558 v.s. 0.556).
These results suggest that ConvRerank can provide benefits regardless of the first-stage environments and improve effectiveness even when adopting 
non-neural first-stage retrieval.

\subsubsection{Model Sizes.}
To examine the impact of model size on the performance of ConvRerank, 
we fine-tine ConvRerank on T5-large and T5-3B\footnote{We only fine-tune the model on T5-3B for 2 epochs due to high computational costs.} with the same procedure and inference setups. 
As observed from Table~\ref{tab:ablation-b}, our ConvRerank benefits more from scaling model size compared to the monoT5 re-ranker (with T5-rewrite).
We hypothesize that the T5-base rewriter bounds the re-ranking effectiveness. 
Thus, to attest to the effectiveness of the multi-stage pipeline (monoT5 w/T5 rewrite), 
we should scale sizes of both the re-ranker and re-writer, which potentially increases query latency.
In contrast, ConvRerank is a single model and does not suffer from this issue, making it an advantageous choice for ConvSearch.

\begin{table}
    \centering
    \caption{Scaling up the model sizes.}
    \vspace{-0.4cm}
    \resizebox{.40\textwidth}{!}{
    \begin{tabular}{llcccc}
    \toprule
         & &
         \multicolumn{2}{l}{CAsT'19 Eval} &
         \multicolumn{2}{l}{CAsT'20 Eval} \\
         \cmidrule(lr){3-4}
         \cmidrule(lr){5-6}
         Re-ranking & Size &
         \multicolumn{2}{l}{nDCG@3 / 100} &  
         \multicolumn{2}{l}{nDCG@3 / 100} \\
    \midrule
         monoT5 (w/T5-rewrite)& \multirow{2}{*}{large} & 
         \multicolumn{2}{l}{0.534 / 0.589} & 
         \multicolumn{2}{l}{0.449 / 0.531}  \\ 
         ConvRerank &  & 
         \multicolumn{2}{l}{{\bf 0.572} / {\bf 0.610}} & 
         \multicolumn{2}{l}{{\bf 0.487} / {\bf 0.550}}  \\ 
         \midrule  
         monoT5  (w/T5-rewrite) & \multirow{2}{*}{3B} &
         \multicolumn{2}{l}{0.534 / 0.592} & 
         \multicolumn{2}{l}{0.470 / 0.545}  \\ 
         ConvRerank & & 
         \multicolumn{2}{l}{{\bf 0.583} / {\bf 0.618}} & 
         \multicolumn{2}{l}{{\bf 0.496} / {\bf 0.562}}  \\
    \bottomrule
    \end{tabular}
    }
    \label{tab:ablation-b}
\end{table}

\section{Conclusion}
We present a novel approach for conversational passage re-ranking, which includes a pseudo-labeling method and our proposed ConvRerank model.
Particularly, we design a view-ensemble method to synthesize high-quality pseudo labels that are then used to fine-tune ConvRerank. Moreover, 
ConvRerank followed by conversational dense retriever as the first-stage retrieval has demonstrated superior performance over other baseline systems on the TREC CAsT datasets in terms of both effectiveness and re-ranking latency.
Moving forward, we plan to strengthen dependencies between the retriever and re-ranker, for instance, by 
(1) implementing a co-training framework~\cite{qu2020openretrieval}, and 
(2) adopting first-stage candidate pruning techniques~\cite{li2022certified}, to improve effectiveness and efficiency.
\begin{acks}
We would like to thank the support of Cloud TPUs from Google’s
TPU Research Cloud (TRC).
\end{acks}

\bibliographystyle{ACM-Reference-Format}
\balance
\bibliography{paper}



\end{document}